\documentclass[useAMS,usenatbib]{mnras}
\usepackage{tabularx}
\usepackage{url}
\usepackage{graphicx}
\usepackage[usenames]{color}

\usepackage[T1]{fontenc}
\usepackage{ae,aecompl}
\usepackage{booktabs}
\def\unit #1{\,{\rm #1}}

\newcommand\kms{\rm \,\unit{km\,s^{-1}}}

\newcommand\cmsqi{\rm \,\unit{cm^{-2}}}

\newcommand\kev{\rm \,\unit{keV}}

\newcommand\ergs{\rm \,\unit{erg\,s^{-1}}}

\newcommand\funit{\rm \,erg\,cm^{-2}\,s^{-1}}

\newcommand\lunit{\rm \,erg \,s^{-1}}

\newcommand\xiunit{\rm \,erg\,cm\,s^{-1}}

\newcommand\msol{M_{\odot}}

\newcommand\ks{\, \rm ks}

\newcommand\dc{\, \Delta\chi^2}

\newcommand\cd{\,\rm \chi^2/dof}

\newcommand\mpc{\unit{Mpc}}

\newcommand\ev{\unit{\, eV}}

\def\ltsim{\mathrel{\hbox{\rlap{\hbox{\lower3pt\hbox{$\sim$}}}\hbox{\raise2pt\hbox{$<$}}}}}
\def\gtrsim{\mathrel{\hbox{\rlap{\hbox{\lower3pt\hbox{$\sim$}}}\hbox{\raise2pt\hbox{$>$}}}}}

\newcommand\asca{{\it ASCA}}
\newcommand\sax{{\it BeppoSAX}}

\newcommand\suzaku{{\it Suzaku}}

\newcommand\swift{{\it Swift}}
\newcommand\xmm{{\it XMM-Newton}}
\newcommand\nustar{{\it NuSTAR}}

\newcommand\atca{{\it ATCA}}
\def\pasp{PASP} 
\def\apj{ApJ} \def\mnras{MNRAS}
\def\aap{A\&A} \def\apjl{ApJ} \def\aj{aj} 
 \def\apjs{ApJS}  \def\pasj{PASJ}
 \def\ssr{SSRv}  \def\araa{ARAA}

\title{Accretion disk/corona emission from a  radio-loud narrow line Seyfert 1 galaxy PKS~0558--504}

\author [Ghosh, Dewangan \& Raychaudhuri] {R. Ghosh$^{1}$\thanks{Email: riteshghosh.rs@visva-bharati.ac.in}, 
G.\ C.\ Dewangan$^{2}$\thanks{Email: gulabd@iucaa.in} , B. Raychaudhuri$^{1}$\thanks{Email: biplabphy@visva-bharati.ac.in} \\
$^{1}$ Visva-Bharati University, Santiniketan, India \\ 
$^{2}$ Inter-University Centre for Astronomy and Astrophysics (IUCAA), Pune, India}

\begin{document}
\maketitle


\begin{abstract}   
Approximately $10-20\%$ of Active Galactic Nuclei are known to eject powerful jets from the innermost regions. There is very little observational evidence if the jets are powered by
spinning black holes and if the accretion disks extend to the innermost regions in radio-loud AGN. Here we study the soft X-ray excess, the hard X-ray spectrum and the optical/UV emission from the radio-loud narrow-line Seyfert 1 galaxy PKS~0558--504 using \suzaku{} and \swift{} observations.
The broadband X-ray continuum of PKS~0558--504 consists of a soft X-ray excess emission below $2\kev$ that is well described by a blackbody ($kT\sim 0.13\kev$) and high energy emission that is well described by a thermal Comptonisation (compps) model with $kT_e\sim 250\kev$, optical depth $\tau\sim 0.05$ (spherical corona) or $kT_e\sim90\kev$, $\tau\sim0.5$ (slab corona). The Comptonising corona in PKS~0558--504 is likely hotter than in radio-quiet Seyferts such as IC~4329A and Swift~J2127.4+5654. The observed soft X-ray excess can be modeled as blurred reflection from an ionised accretion disk or optically thick thermal Comptonisation in a low temperature plasma. Both the soft X-ray excess emission when interpreted as the blurred reflection and the optical/UV to soft X-ray emission interpreted as intrinsic disk Comptonised emission implies spinning ($a>0.6$) black hole. These results suggest that disk truncation at large radii and retrograde black hole spin both are unlikely to be the necessary conditions  for launching the jets. 

\end{abstract}


\section{Introduction}
Active galactic nuclei (AGNs) are known to exhibit one of the most energetic phenomena in the universe. A small subset ($\sim10-20\%$) of AGNs are radio-loud with powerful radio emission from relativistic jets \citep{1989AJ.....98.1195K,1995PASP..107..803U,2002AJ....124.2364I}. 
The physical processes that distinguish between radio-loud and radio-quiet AGNs are still unknown. The jet formation and the radio emission is generally assumed to be connected with the presence and structure of an accretion disk, but the connection between the formation of the jet and the central engine has been a topic under debate for a long time~\citep[see, e.g.,][]{2001bhbg.conf..206C}.
It is therefore important to study the innermost disk/corona regions in radio-loud AGN that can unravel the mystery of the jet launching process. 

Over the past four decades, though the black hole accretion theory has
 gained confidence, the understanding of the jet-disk coupling  are still poorly understood. The variability properties of hard-X-rays from AGNs indicate that they are produced from the innermost regions of the accreting material. The presence of the Compton hump above $\gtrsim 8
\kev$ and $Fe K_{\alpha}$ line are naturally interpreted as evidence of reprocessing of the accreting material. But over the last few decades, observations of majority radio-quiet populations and minority radio-loud sources by \asca{} and \sax{} indicates that
radio-loud AGNs may have weaker hard X-ray reprocessing features than radio-quiet type 1 counterparts \citep{1998ApJ...505..577E,1999A&A...343...33G}. This could be due to an accretion disk that is truncated in the inner regions and changes to a hot optically thin flow before reaching the black hole. 
 Recent observations also suggest that radio-loud quasars, on average, have lower Eddington ratios compared to radio quiet quasars~\citep{2007ApJ...658..815S}.
%
Alternatively the weak reflection features may indicate the presence of an ionized untruncated disk provided the accretion rate is a larger fraction of Eddington \citep{2002MNRAS.332L..45B}. In order to make progress in the understanding of the formation of jets, it is crucial to study the conditions in the innermost regions, including the inner accretion disk and the hot corona in radio-loud AGNs. Till date, there is very little observational evidence if the spinning black holes are required for the formation and launching of jets and/or the accretion disk is truncated in radio-loud AGNs. Recent multi-epoch analysis of \xmm{} and \suzaku{} observation of radio loud Seyfert-1 $3C~120$ indicates the possibility of disk-disruption or jet cycles in the innermost region~\citep{2013ApJ...772...83L}. Again radio interferometry observations at 1.3 millimeters of radio loud
elliptical galaxy $M87$ seem to spatially resolve the base of jet production in the accretion disk around $\sim 3 r_{g}$ and also suggest a prograde accretion disk around a spinning black hole~\citep{2012Sci...338..355D}. These possible scenarios require detailed analysis of the broadband spectral features of a radio-loud AGN. 
Radio-loud narrow-line Seyfert 1 galaxies (\citealt{2006AJ....132..531K}) with high accretion rates relative to the Eddington rate and lower masses are best suited for this purpose. 

PKS~0558--504 (z = 0.137) is an radio-loud NLS1 with optical spectrum similar to 
that of radio-quiet narrow-line Seyfert 1 galaxies~\citep{1986BAAS...18..915R,2006AJ....132..531K}.  
Probing the radio Sky at Twenty-centimeters of the Very Large Array~\citep{2012ApJ...760...41D} revealed the two sided radio structure on kpc scales and one 
sided jet morphology on pc scale.
Multiwavelength observations using \xmm{}, \swift{} and \atca{},  \citet{2010ApJ...717.1243G}  constrained the BH mass, $M_{BH}\approx 2.5\times10^8\msol$, and confirmed that PKS~0558--504 is accreting at or above the super-eddington rate  earlier found by \citealt{2007ApJ...656..691G}. \citet{2010ApJ...717.1243G} also found that the spectral energy distribution(SED) of PKS~0558--504 is dominated by the optical-UV emission and the jet emission do not dominate beyond the radio band. Previously \citet{2001MNRAS.323..506B} used blurred reflection model to fit the \asca{} data and although the reflection fraction was not well constrained, a possible weak $Fe K_{\alpha}$ line was found and the emission was consistent with occurring within the $10 r_{g}$. \citet{2010A&A...510A..65P} used similar models, but found a poor fit and $Fe K_{\alpha}$ was found to be weak with an equivalent width(EW)$\sim 20 \ev$ for $90\%$ confidence level. Later \citet{2013MNRAS.428.2901W} found a good fit with the blurred reflection model and the best fit BH spin $a\sim 0.9$. These characteristics make PKS~0558--504 an interesting as well as an ideal radio-loud AGN to probe the intrinsic emission spectra from inner regions. 
In this paper, we study the  nuclear broadband  emission of PKS~0558--504 using \suzaku{} and \swift{} data.

 The paper is organized as follows. First, we describe the data sets used in this work and briefly discuss the data reduction techniques in Section 2. In Section 3 we first perform a preliminary, basic investigation of spectral shape of the individual data epochs and later a multi-epoch analysis is used to investigate the nature of the X-ray spectrum. Finally, Section 4 contains a discussion of the results.
We assume a cosmology with $H_{0} = 71\kms \mpc^{-1}, \Omega_{\Lambda} = 0.73$ and $\Omega_{M} = 0.27$  to calculate the distance.

\section{Observation and Data Reduction}

\subsection{\bf \suzaku}
The five \suzaku{} observations of PKS~0558--504 were performed using X-ray
Imaging Spectrometer (XIS)~\citep{2007PASJ...59S..23K} and Hard X-ray Detector (HXD)~\citep{2007PASJ...59S..35T}, 
spanning a period from January 17 to 21, 2007. The exposure time for each
observation is  $\sim 20 \ks$. These observations are summarized
in Table \ref{Observation Table}. The three XIS (XIS0, XIS1 and XIS3)
CCD cameras cover the energy range $0.2-12.0 \kev$ and the HXD/PIN
covers the high energy $10-70\kev$ band. For the first three observations (observation IDs 701011010, 701011020 and 701011030, identified by ``obs-1'',
``obs-2'', etc.), the XIS data were obtained in both the $3 \times 3$
and $5 \times 5$ data modes, while for the last two observations (obs-4 and obs-5), the XIS data were observed only in the $3 \times 3$ data mode.

We used {\tt HEASOFT, version 6.16} software and the recent calibration data to process the \suzaku{} data.  We followed the \suzaku{} ABC
Guide\footnote{http://heasarc.gsfc.nasa.gov/docs/suzaku/analysis/abc/}.
We reprocessed and cleaned the unfiltered event files and created the
cleaned event files using the {\sc aepipeline} tool.  In all observations,
for both the XIS0 and XIS3(front-illuminated CCD) and for XIS1(back-illuminated CCD), we extracted the source spectra for each observation from the filtered event lists using a $250{\rm~arcsec}$ circular region centered on the source position. We also extracted the corresponding Background spectral data using
multiple circular region of $120{\rm~arcsec}$ radii, excluding the source region. We generated the ancillary response and the redistribution matrix
files for each XIS spectral data by using the {\tt xissimarfgen} and
{\tt xisrmfgen} tools, respectively.

We also extracted the hard X-ray spectral data using the {\sc
  hxdpinxbpi} tool from the PIN cleaned events and the pseudo event lists generated by the {\sc aepipeline} tool. For the non-imaging HXD/PIN data, the background estimation requires both non X-ray instrumental background (NXB) and the cosmic X-ray background (CXB). We used the appropriate tuned background files provided by the \suzaku{} team and available at the {\sc HEASARC}
website~\footnote{http://heasarc.gsfc.nasa.gov/docs/suzaku/analysis/pinbgd.html}. We grouped the XIS spectral data to a minimum of $100$ counts in each energy bin. We also grouped the PIN data to produce $\sim $ 60 energy
bins with more than 20 counts per bin in the source spectra.

\subsection{\bf \swift}
PKS0558--504 has been monitored 
with the \swift{} mission~\citep{2004ApJ...611.1005G} between 2008 September 7 and 2010 March 30. During the first 10 days the source was observed on a daily basis and the optical/UV emission was weakly variable with  fractional variability $F_{var} \ltsim 0-4\%$ \citep{2010ApJ...717.1243G}. Here we have used the third observation performed on September 9, 2008. Our purpose here is to derive optical-to-hard X-ray spectrum of PKS~0558--504. The UVOT
~\citep{2005SSRv..120...95R} instrument observed the source PKS~0558--504
 in all six filters. We calculated the source and background rates  from
 the co-added image files and using circular regions of radius $5^{''}$ 
for the source and $20^{''}$ for the background. We used the UVOT2PHA tool
 to create the source and background pha files, and used the response files
 provided by the \swift{} team. 

The Swift optical points are not simultaneous with the \suzaku{} XIS observation and we compared our results, with the \swift{} XRT instrument~\citep{2005SSRv..120..165B}. 
The Swift XRT observation was performed in Windowed Timing mode in order to avoid the effects of pile-up. We analyzed the XRT data from same observations mentioned above, with standard procedures using {\tt xrtpipeline}. The HEASOFT package version 6.16 and recent calibration database (CALDB) is used for filtering, and screening of the data. The source and background regions were selected in boxes 40 pixels long. We use the standard grade selections of 0-2 for the Window timing mode. Source photons for the light curve and spectra were extracted with XSELECT. The auxiliary response files (ARFs) were created using {\tt xrtmkarf} and the response matrix $swxwt0to2s6_20010101v015.rmf$. We bin the data using grppha
to have at least 20 counts per bin.

\begin{figure*}
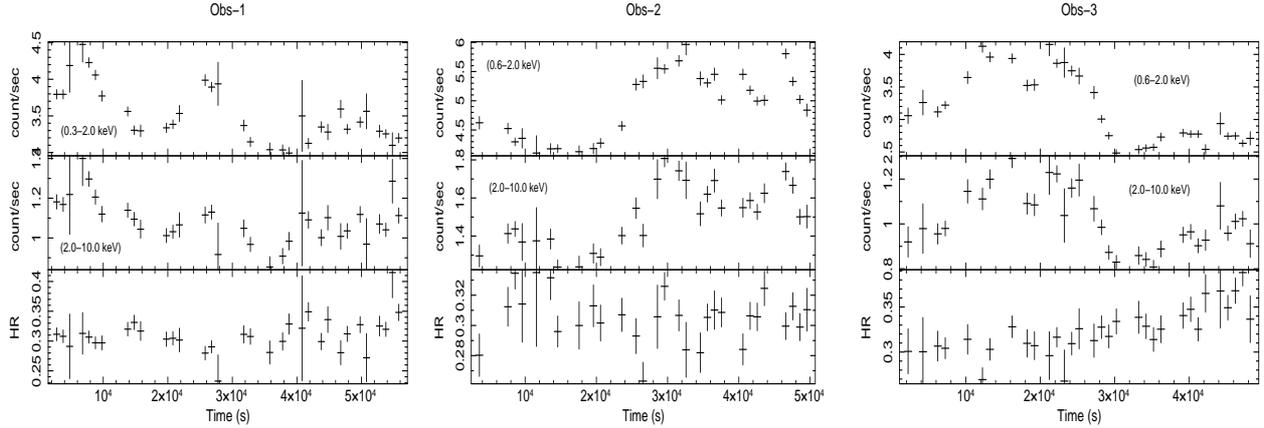


  \includegraphics[width=5.7cm,height=5.5cm,angle=-90]{Hardness_ratio_xis_bin1000s_obs1.ps}
  \includegraphics[width=5.7cm,height=5.5cm,angle=-90]{Hardness_ratio_xis_bin1000s_obs2.ps}
  \includegraphics[width=5.7cm,height=5.5cm,angle=-90]{Hardness_ratio_xis_bin1000s_obs3.ps}
  \caption{Background subtracted XIS lightcurve of PKS~0558--504 in the 0.6--2 $\kev$ (Top) and 2.0--10.$\kev$ (Middle) and the hardness ratio(HR) (Bottom) for obs-1(Left), obs-2(Middle) and 0bs-3(Right) respectively. The hardness ratio is almost constant($\sim 0.3$) during these observations.}
\label{Hardness_ratio}
  
\end{figure*}

\begin{table}
\centering
\footnotesize
  \caption{The five \suzaku{} ans \swift{} XRT observations of PKS~0558-504 used in this work.  \label{Observation Table}}
\setlength{\tabcolsep}{2pt}
\begin{tabular}{cccc} \hline\hline
       ObsID 	  & Observation &  Exposure      &   Net count rate  \\
                  & date &        (ks)           &  (${\rm counts~s^{-1}}$)  \\ 
            	  &      &     XIS/HXD         &       XIS0/HXD     \\  \hline 
   
   701011010(obs-1)  &2007-01-17 & 21/19 & $1.54^{+0.01}_{-0.01}/0.047^{+0.006}_{-0.006}$ \\
   701011020(obs-2)  &2007-01-18 & 19/18 & $2.15^{+0.01}_{-0.01}/0.038^{+0.007}_{-0.007}$ \\
   701011030(obs-3)  &2007-01-19 & 21/19 & $1.35^{+0.01}_{-0.01}/0.024^{+0.006}_{-0.006}$ \\ 
   701011040(obs-4)  &2007-01-20 & 20/17 & $2.21^{+0.02}_{-0.02}/0.047^{+0.007}_{-0.007}$ \\
   701011050(obs-5)  &2007-01-21 & 20/16 & $2.24^{+0.01}_{-0.01}/0.032^{+0.007}_{-0.007}$  \\
   00090020003(XRT)  &2008-09-09 & 2.3   & $0.98^{+0.02}_{-0.02}$ \\\hline 
\end{tabular}  
\end{table}

\section{Spectral analysis and Results}

We used {\sc XSPEC} version 12.8.2 to analyse all the spectral data. We used 
$\chi^{2}$ statistics and quote the errors on the best-fit parameters at 
the $90\%$ confidence level. Galactic absorption due to neutral Hydrogen with a column density of
 ($N_H = 4.4\times10^{20}\cmsqi$; \citet{1990ARA&A..28..215D}) is included in all spectral
models using the {\tt Wabs} model. For all five \suzaku{} observations, we added the spectral data obtained with the XIS0 and XIS3 cameras, the front-illuminated CCDs, using the {\sc
  addascaspec} tool to enhance the signal to noise ratio. We also checked the XIS spectra to find any discrepancy between the XIS0+XIS3 and XIS1 datasets. We discarded the $1.7-2.0\kev$ energy range for all five observations as the XIS datasets show calibration uncertainties in above mentioned spectral range. We also ignored bad channels from our spectral analysis.

 We begin with spectral analysis of the \suzaku{} data to check the presence of possible spectral variability during five observations and as Table~\ref{Observation Table} shows large variations in count rate between observations 1, 2 and 3, the soft band ($0.6-2.0\kev$), hard band ($2.0-10.0\kev$) lightcurves and corresponding hardness ratios are plotted in Fig.~\ref{Hardness_ratio} using XIS data from these observations and it indicates against any possible spectral variability with the hardness ratio having a constant value around 0.3.  

\begin{figure}

  \includegraphics[width=5.8cm,angle=-90]{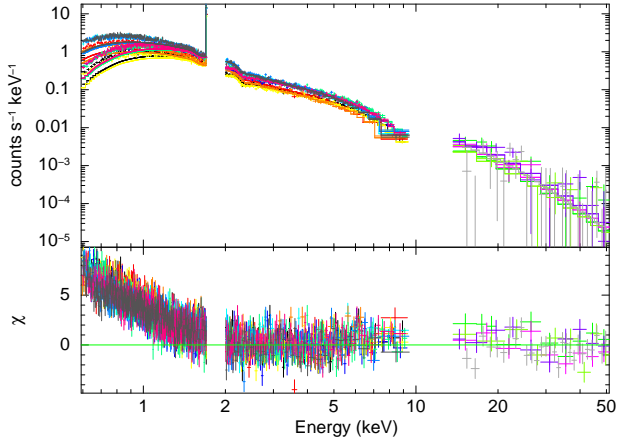}
  
  \caption{The joint XIS+PIN spectral data from all obs, the absorbed powerlaw
    model and the deviation of observed data from the powerlaw model
    showing soft excess.\label{Residuals_absorbed_powerlaw}}
  
\end{figure}

 We then performed a simultaneous spectral fitting of XIS0+XIS3, XIS1 and HXD/PIN data for each observation. We used a simple powerlaw model modified with the Galactic absorption and also
multiplied a constant model to account for the relative normalizations of different instruments.  We fixed the constant to 1 for XIS0+XIS3 combined data and allowed it to vary for XIS1. For the HXD/PIN data, we fixed the relative normalization in 1.16 as the observations were performed at the XIS nominal position. We fitted the absorbed powerlaw model in the $2-10\kev$ band and
then extrapolated to the lower energies down to $0.6\kev$ and to the high energy of the $15-50\kev$ PIN band. A prominent soft excess below $2\kev$ along with a possible weak hard excess beyond $10 \kev$ was observed for all five observations. We repeat our analysis with a joint fit of all five observations and in Fig.~\ref{Residuals_absorbed_powerlaw}, we show the \suzaku{} spectral data, the absorbed powerlaw model and the deviations of the observed data from the model.

\begin{figure}

  \includegraphics[width=6cm,angle=-90]{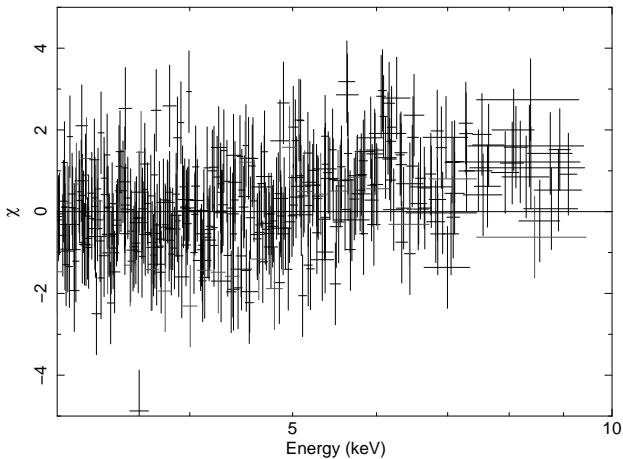}
  
  \caption{Rebinned and zoomed residual at $6 \kev$ derived from a joint fit to five Suzaku observations with absorbed powerlaw and {\tt bbody} \label{residual_at_6kev}}
\end{figure}

Both phenomenological such as single or multiple blackbodies, multicolor disk blackbody and physical models, e.g., blurred reflection from partially ionised medium, thermal Comptonization in an optically thick medium provide  statistically good fit to the soft-excess. In case of PKS~0558--504 in particular,~\cite{2001A&A...365L.122O} used three black body components to fit
this excess. \cite{2004A&A...415..959B} tried Comptonisation component and~\cite{2008PASJ...60..487H} used multicolour disc black body model. We did not use any complex absorption models as no discrete absorption features were detected in the high resolution RGS spectra obtained with XMM-Newton~\citep{2010A&A...518A..28P}. We pursued a multi-epoch fit of the five \suzaku{} data sets where all five spectra are  fitted simultaneously, and thus increasing the statistical significance of the fits. 

\subsection{Phenomenological models}
We start with the absorbed powerlaw model and it again provides a poor fit with $\chi^{2}=8663.9$ for 4060 degrees of freedom(dof). The addition of a simple {\tt bbody} to the absorbed powerlaw model improved the fit significantly with $kT\sim 0.13 \kev$ and $\Gamma \sim 2.2$ 
(with $\dc \sim -4330$ to $\chi^{2}$/dof=4364/4050). A hint of  residual at the energy of the $Fe K_{\alpha}$ line is observed (see Fig~\ref{residual_at_6kev}). To check it further, we first introduced a narrow Gaussian line profile (line width fixed at $0.01 \kev$) to the absorbed powerlaw and bbody best fit. It improves the fit by $\dc \sim -40$ to $\chi^{2}$/dof=4326/4048 indicating a cold distant reflection. We then add a broad Gaussian to the previous best fit and although it improves the fit, we were unable to constrain the line width. Assuming a broad Gaussian with a line width $0.1 \kev$ the improvement in statistics was $\dc \sim 14$ from the narrow Gaussian best fit indicating a blurred reflection like feature. In order to check the significance of a relativistically broadened component we removed the gaussian components and replaced with {\tt relline}~\citep{2010MNRAS.409.1534D} model and found an improvement of 63 in $\chi^{2}$ value over powerlaw plus bbody best fit with six additional parameters and the rest-frame energy is 6.73 keV which indicate towards a helium like Iron. Although the results indicate towards a reflection dominated spectra we note that we were unable to constrain the best fit spin parameter and apart from the soft excess below $2.0 \kev$ energy band, data provides no direct evidence for relativistic reflection.

\subsection{Reflection models}
To further check the possibility of a reflection dominated spectra
we then removed the {\tt relline} model and replaced with {\tt pexmon} model which predicts a cold, distant reflection along with self-consistently generated Fe emission lines. We used powerlaw for the continuum, {\tt bbody} for soft excess and set {\tt pexmon} to generate reflection only, by setting the scaling factor for reflection(R) negative. In XSPEC the model reads as {\tt wabs$\times$(po+bbody+pexmon)}. It provided an equally good fit; ($\chi^{2}$/dof=4325/4049), with a best fit relative reflection of R=$-0.18^{+0.05}_{-0.05}$. Initially Fe abundance was fixed to the solar value; making it a free parameter further improves the fit to $\chi^{2}$/dof=4223/4048 with best  fit Fe abundance value of $0.08^{+0.05}_{-0.04}$ and R=$-0.64^{+0.08}_{-0.08}$. This result confirms the presence of a strong, smooth, soft excess along with reflection features.

\begin{figure}
  \includegraphics[width=6.5cm,height=8cm,angle=-90]{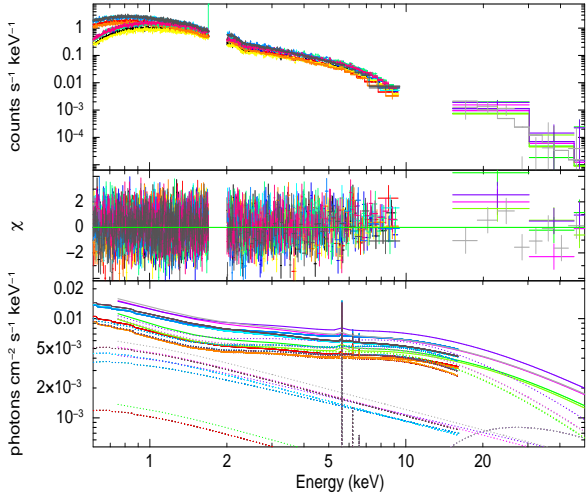}
 
  \caption{The observed XIS0+3 and PIN spectral data, the best fit blurred reflection model(upper panel), Deviation of the observed data from the best fit model(middle panel) and spectrum model(lower panel) are plotted. The model reads as: {\tt wabs$\times$(po+relxill+pexmon)}.}
  \label{relxill_fit_plot}
  
\end{figure}

\begin{figure*}
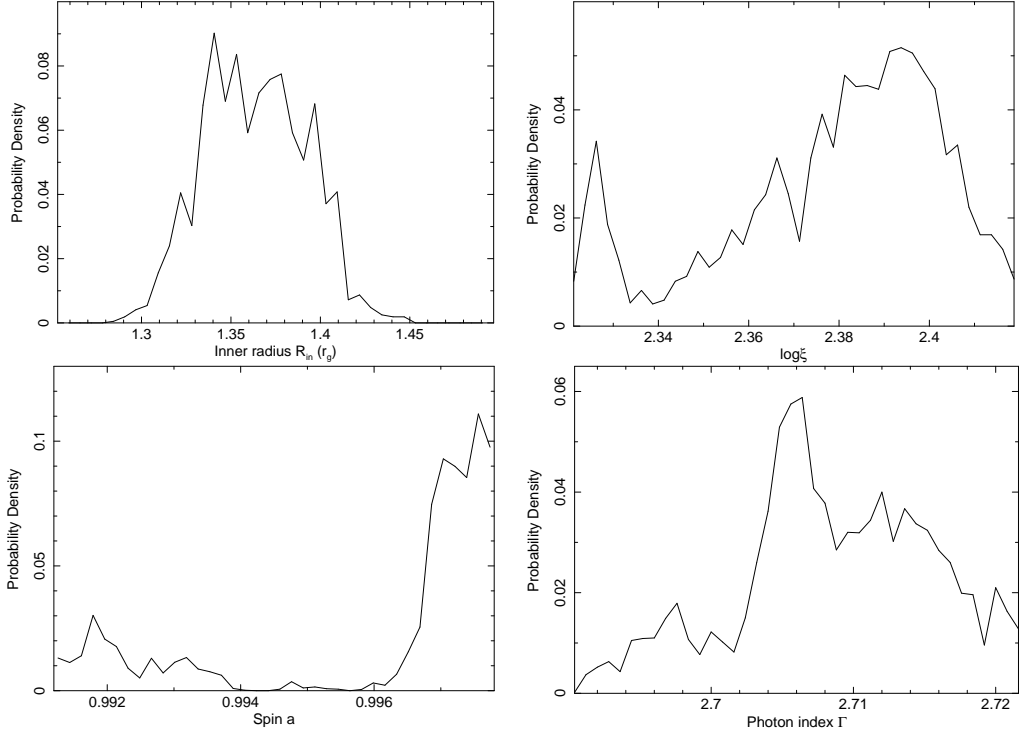


  \includegraphics[width=4.8cm,angle=-90]{Probability_distribution_new_Rin_relxill.ps}
  \includegraphics[width=4.8cm,angle=-90]{Probability_distribution_new_logxi_relxill.ps}
  \includegraphics[width=4.8cm,angle=-90]{Probability_distribution_new_spin_relxill.ps}
  \includegraphics[width=4.8cm,angle=-90]{Probability_distribution_new_gamma_relxill.ps}
\caption{MCMC results:{\it Left top:} Probability distribution for $R_{in}$ {\it Left bottom:} Probability
 distribution for $a$ {\it Right top:} Probability distribution for $log \xi$ and {\it Right bottom:} Probability distribution for $\Gamma $ for best fit {\tt relxill} plus {\tt pexmon} model with powerlaw.}
\label{probability_relxill}
  
\end{figure*}

{\tt Relxill}~\citep{2014ApJ...782...76G} is another physical model which predicts $Fe K_{\alpha}$ line along with reflection continuum i.e., both soft excess and Compton hump. 
We removed the {\tt bbody} and {\tt pexmon} models from the previous best fit and have used the latest {\tt Relxill} version which includes both the {\tt XILLVER} reflection code~\citep{2010ApJ...718..695G} and  the {\tt RELLINE} code~\citep{2010MNRAS.409.1534D}. This model for each point on the disk, assumes a proper xillver-reflection spectrum for all relativistically calculated emission angles where {\tt XILLVER} calculates the reflected spectrum emerging from the surface of an X-ray illuminated accretion disk by simultaneously solving the equations of radiative 
transfer, energy balance, and ionization equilibrium in a Compton-thick, plane parallel medium. In XSPEC the model reads as {\tt wabs$\times$(po+relxill)}. It produced a good fit for the multi-epoch fit of the 5 \suzaku{} data with $\chi^{2}$/dof=4349/4051. The addition of the {\tt pexmon} model improves the fit by $\dc \sim 20$ to $\chi^{2}$/dof=4326/4051. The best fit parameters are listed in the Table~\ref{relxill fit}. The data set, best fit models and deviations of the data for {\tt relxill} model is shown in Figure~\ref{relxill_fit_plot}. The iron abundance($\sim 0.89$) being consistent with Solar value, the best fit parameter values, e.g. R $\sim 4.31$, log $\xi = 2.37$, a $\sim 0.99$ and $R_{in}(r_{g}) = 1.32^{+0.08}_{-0.08}$ indicates that 
the inner radius of the disk tends to lie well within the $6 r_{g}$ which is the last stable orbit around a non-rotating Schwarzchild black hole. These results indicate towards a rapidly rotating black hole, although we note that both the high spin (a $\sim 0.99$) and high reflection parameter ($R \sim 4.3$) is required by the model to fit the spectra due to the presence of prominent soft excess and possible broad Fe emission line. The results of the Monte Carlo Markov Chain (MCMC) analysis for selected parameters are done to make sure that the best fit parameters are not stuck in any local minima and the fit yielded a similar probability density and shown in Fig.~\ref{probability_relxill}. 


\begin{table}
\footnotesize
\centering
  \caption{Best fit parameters for \suzaku{} observations of PKS~0558-504 for the model {\tt wabs$\times$(po+relxill+pexmon)} \label{relxill fit}}

  \begin{tabular}{lll} \hline
Component  & parameter                &\\ \hline
Gal. abs.  & $N_{H}(10^{20}\rm cm^{-2})$ & $ 4.4$ (f)         \\ 

powerlaw   & $ \Gamma $               & $2.72^{+0.01}_{-0.01}$ \\
           & $ n_{pl}(10^{-3}) ^a$    & $1.14^{+0.09}_{-0.10}$ \\
relxill    &  $A_{Fe}$                & $0.89^{+0.11}_{-0.13}$ \\ 
           &  $log\xi (\xiunit)$      & $2.37^{+0.04}_{-0.03}$ \\ 
           & $ \Gamma $               & $2.72^{+0.01}_{-0.01}$ \\
           &  $n_{rel}(10^{-3})^a$    & $3.11^{+0.19}_{-0.18}$ \\
           &   $ q$                   & $ >7.48$ \\
           &   $ a$                   & $ >0.995$ \\
           &   $R(refl frac) $        & $4.31^{+0.37}_{-0.16}$ \\
           &   $ R_{in}(r_{g})$       & $1.32^{+0.11}_{-0.03}$ \\
           &   $ R_{br}(r_{g})$       & $4.51^{+2.75}_{-0.32}$ \\
           &   $ R_{out}(r_{g})$      & $400$ (f)              \\
           &   $i(degree) $           & $37.68^{+8.04}_{-6.68}$ \\
pexmonn    &   $R $                   & $-0.76^{+0.21}_{-0.30}$ \\ \hline
           & $\cd $                   & $4326/4051$           \\\hline 
\end{tabular} \\ 
Notes: (f) indicates a frozen parameter.
(a) $n_{pl}$ and $n_{rel}$ reperesent normalization to respective model component; where $n_{pl}$ has the unit as photons $\kev^{-1} cm^{-2}s^{-1}$. 
\end{table}

 \subsection{Comptonisation models}

\subsubsection{Double Comptonisation model}

\begin{figure}

  \includegraphics[width=6.5cm,height=8cm,angle=-90]{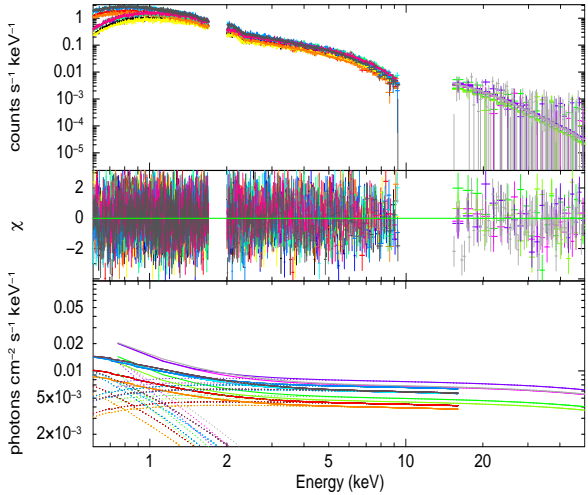}
  \caption{The observed XIS0+3 and PIN spectral data, the best fit double Comptonisation model(upper panel), Deviation of the observed data from the best fit model(middle panel) and spectrum model(lower panel) are plotted. The model reads as: {\tt wabs$\times$(nthcomp+nthcomp)}.}
\label{nthcomp fig}
  
\end{figure}

To investigate the issue of low-temperature Comptonisation, we first considered a double Comptonized model using the model $nthcomp$~\citep{1996MNRAS.283..193Z,1999MNRAS.309..561Z}. The common interpretation is that the hot corona is the inner part of the accretion flow while the warm corona could be the upper layer of the outer optically thick accretion disk. So one of the nthcomp models should describe the soft excess, and we varied the $kT_{e}$ for this component, while we fixed the $kT_{bb}$ to expected inner disk temperature of $10 \ev$. The value was calculated assuming an $M_{BH}$ mass of $2.5\times10^{8}\msol$, accretion rate of 0.5, relative to the Eddington rate and r = 6$R_{S}$. For the second nthcomp, we made the parameter $kT_{bb}$ free as the soft excess is supposed to be the seed photons for the nthcomp. In XSPEC the model reads as {\tt wabs$\times$(nthcomp+nthcomp)}
 This double Comptonized model provides a good fit with $\chi^{2}$/dof=4238/4052, see Table~\ref{nthcomp fit}. Addition of $pexmon$ model does not improve the fit significantly as $\dc \sim -3$ is observed with similar best fit parameters. The data, folded model and the deviation of the model is plotted and shown in Fig.~\ref{nthcomp fig}. We note that the fit does not explicitly require a spinning black hole and indicates towards the presence of other possible values of a. However, the best-fit parameters are based on \suzaku{} X-ray data only and this model does not explain the origin of the UV emission from the disk.

\begin{table}
\footnotesize
\centering
  \caption{Best fit parameters for \suzaku{} observations of PKS~0558-504 for 
absorbed double Comptonization model.\label{nthcomp fit}}

  \begin{tabular}{lll} \hline
Component    & parameter                &\\ \hline
Gal. abs.    & $N_{H}(10^{20}\rm cm^{-2})$ & $ 4.4$ (f)         \\ 

nthcomp (1)  & $ \Gamma $               & $3.71^{+0.17}_{-0.12}$ \\
             & $ n(10^{-3}) $           & $3.52^{+0.72}_{-0.48}$ \\
             &  $kT_{e}(\kev)$          & $ >0.59$ \\ 
             &  $kT_{bb}(\kev)$         & $0.01$ (f)\\ 
nthcomp (2)  & $ \Gamma $               & $2.05^{+0.05}_{-0.06}$ \\
             &  $n(10^{-3})$            & $4.78^{+0.48}_{-0.71}$ \\
	     &  $kT_{e}(\kev)$          & $300$ (f) \\ 
             &  $kT_{bb}(\kev)$         & $0.14^{+0.04}_{-0.07}$ \\ \hline
           & $\cd $                     & $4238/4052$           \\\hline 
\end{tabular} \\ 
Notes: (f) indicates a frozen parameter.
\end{table}

\subsubsection{The compPS model}

The broadband spectral datasets provide an opportunity to test geometry of the hot corona. We therefore used the {\tt compps} model which predicts the Comptonisation spectra for different geometries using an exact numerical solution of the radiative transfer equation. The seed photons are assumed to arise from a blackbody with a maximum temperature $9.6\ev$, the characteristic
 temperature of an accretion disk around a $2.5\times 10^{8}\msol$ black hole with the assumption of $\dot{M}$ = 0.5, $r_{in} = 6 r_{g}$ . In addition to the plasma temperature $kT_{e}$, the Compton $y$ parameter, defined as $y = 4\tau(kT_{e}/511 \kev)$ is used as a fit parameter instead of the optical depth $\tau$. The {\tt compps} model can predict spectra for several different geometries of the Comptonising plasma (e.g., sphere, slab, or cylinder). All the geometries produce qualitatively similar results, so we focus here on the fits from the spherical and slab geometry. 

\begin{figure}

  \includegraphics[width=5.5cm,angle=-90]{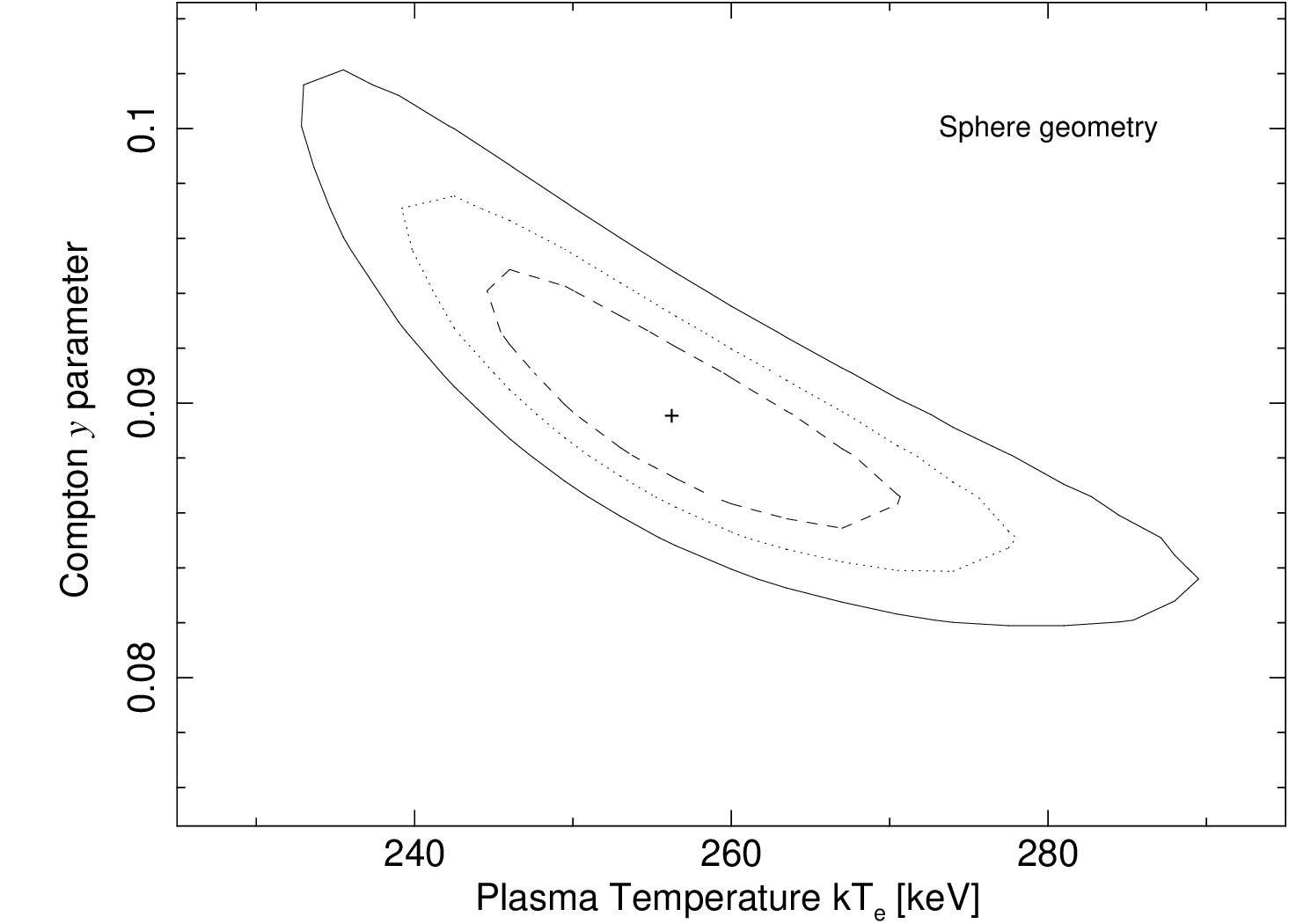}
  \caption{Contour plot of $kT_{e}$ vs $\tau{}$ for compPS model best fit of five \suzaku{} data. Dashed, dotted and solid lines represent 68\%, 90\% and 95\% confidence contours respectively.}
\label{compps sphere contour}
  
\end{figure}

\begin{figure}

  \includegraphics[width=5.5cm,angle=-90]{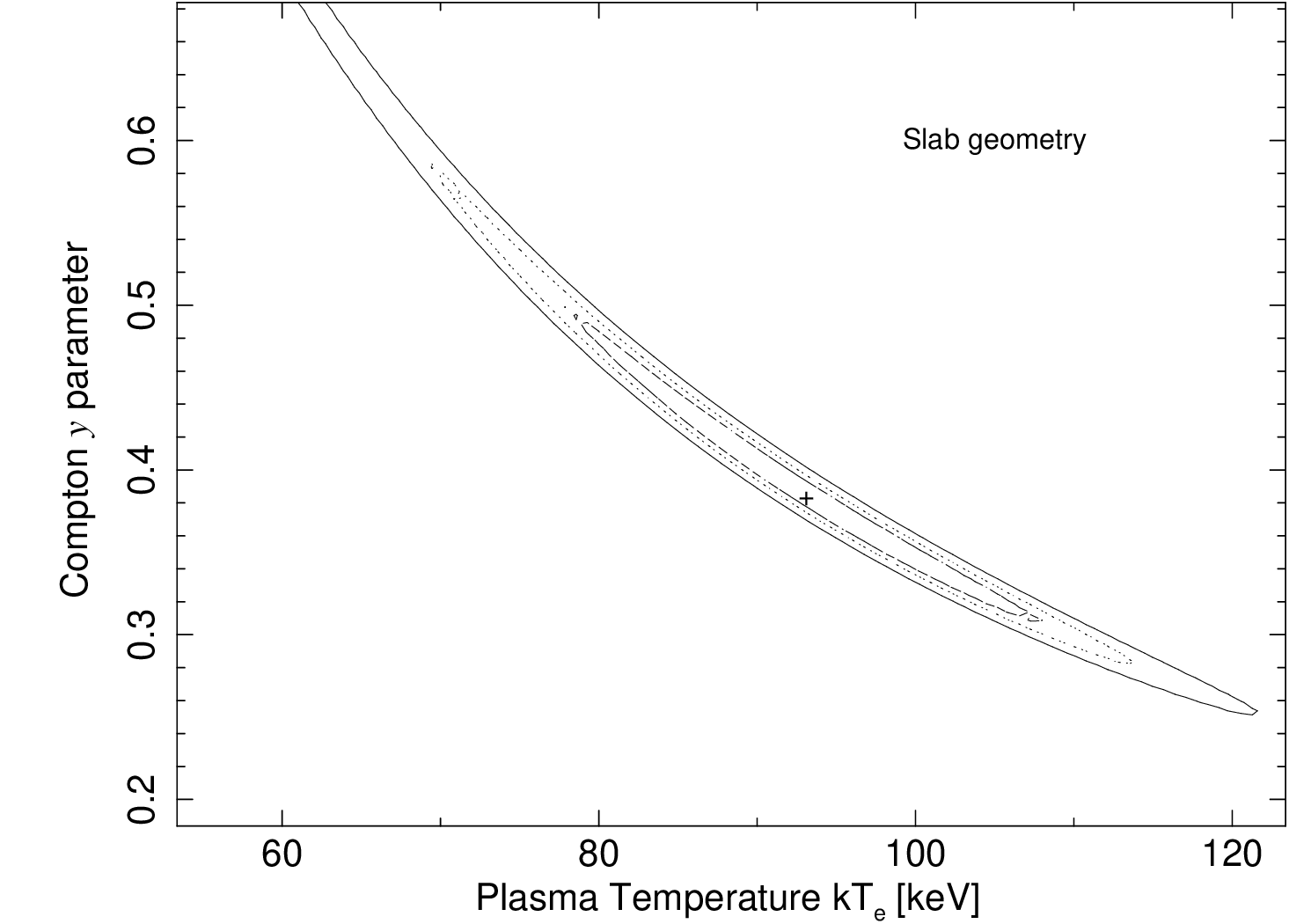}
  \caption{Contour plot of $kT_{e}$ vs $\tau{}$ for compPS model best fit of five \suzaku{} data. Dashed, dotted and solid lines represent 68\%, 90\% and 95\% confidence contours respectively.}
\label{compps slab contour}
  
\end{figure}

First, we tried a {\tt bbody} component to account for the soft excess below $0.6\kev$. 
But as we vary the relative reflection parameter({\tt R}), it takes very high values($\sim 10$) and the bbody parameter seems to change, which indicates part of the soft excess is being described by reflection. So to get reasonable values of {\tt R} we ignore the data below $2 \kev$ and remove the {\tt bbody} component. First, we fixed the {\tt R} to zero and observe that the spectral model for both the geometry provides a satisfactory fit to the data with $\chi^{2}$/dof=1672/1541 and $\chi^{2}$/dof=1696/1541 for the sphere and slab geometry respectively. The temperature of the Comptonizing electrons to be $kT_{e} = 256.5^{+16.0}_{-13.6}\kev$ and a Compton $y$ parameter of $0.09^{+0.01}_{-0.01}$ for the spherical geometry and $kT_{e} = 93.3^{+16.2}_{-17.2}\kev$ and a $y$ parameter of $0.38^{+0.13}_{-0.08}$ for the slab geometry (see Table~\ref{compps fit}). These values result in an optical depth of $\tau = 0.05$ and $\tau = 0.5$ for the sphere and slab geometry respectively. Making {\tt R} a free parameter does not improve the fit for spherical geometry, whereas $\dc \sim -20$ is observed for the slab geometry with best fit values of {\tt R} $=$ $0.38^{+0.42}_{-0.36}$ and $1.07^{+0.45}_{-0.41}$ for the sphere and slab geometry respectively. Thus, the broadband spectral data above 2~keV do not clearly rule out the presence of blurred reflection.  The $2-10.0\kev$ flux of PKS~0558--504 determined by both fit is $F_{2-10\kev} \sim 1.2 \times 10^{-11} \funit$ for both the observations, which corresponds to an unobscured luminosity of $L_{2-20\kev} \sim 5.5 \times 10^{44} \lunit$ which is consistent with ~\citet{2007ApJ...656..691G}. Confidence contours for $kT_{e}$ and $y$ are computed using the best fitting compps model for both the geometry to examine the corona properties of PKS~0558--504 more closely (see Fig.~\ref{compps sphere contour} and Fig.~\ref{compps slab contour}).

\begin{table}
\centering

  \caption{Best fit parameters for \suzaku{} observations of PKS~0558-504 for 
absorbed compPS. 
\label{compps fit}}

  \begin{tabular}{llll}\hline
Component  & parameter                &  Sphere(geom)       & Slab(geom) \\ \hline
Gal. abs.  & $N_{H}(10^{20}\rm cm^{-2})$ & $ 4.4$ (f)             & $4.4$   \\ 

compps     & $kT_{e} (\kev)$          & $256.5^{+16.0}_{-13.6}$   & $93.3^{+16.2}_{-17.2}$    \\ 
           & $ y $                    & $0.09^{+0.01}_{-0.01}$    & $0.38^{+0.13}_{-0.08}$     \\
           & $ rel_{refl} $           & $ 0 $ (f)                 & $ 0 $ (f)  \\
           & $ \tau $                 & $ 0.05 $                  & $ 0.5 $\\
           & $ Tdisk(10^{5}K)$        & $ 1.1$ (f)                & $ 1.1 $ (f)\\
           &   $Cos(i) $              & $0.707$ (f)               & $0.707$ (f)\\
           &  $ F_{2-10\kev}^{(a)}$ & $1.2\times10^{-11}$       & $1.2\times10^{-11} $\\ \hline

           & $\cd $                   & $1672/1541$               & $1696/1541$\\\hline 
\end{tabular} \\ 
Notes: (f) indicates a frozen parameter.\\
 $(a)$ The $(2-10)\kev$ flux has the unit as $\funit$.
\end{table}

\subsubsection{Intrinsic disk Comptonisation model}

Finally, we have used the intrinsic disc Comptonisation model 
{\tt optxagnf}. This model combines the disk emission from $r_{out}$
 to $r_{corona}$, producing the big blue bump, thermal Comptonisation in
 a warm optical thick inner disk, from $r_{corona}$ to $r_{in}$ giving rise to the soft X-ray excess and high temperature thermal Comptonisation in an optically thin, hot corona producing the high energy X-ray powerlaw component \citep{2012MNRAS.420.1848D}. {\tt Optxagnf} model assumes the gravitational energy released at each radius to be emitted as colour temperature corrected blackbody down only to $r_{corona}$, and the energy release is distributed into a warm thermalized disk and optically thin, hot corona thus making it a self-consistent model with the assumption that the disc structure changes inside some radius $r_{corona}$. The key aspect of the model is that the 
 luminosity of the soft excess and tail are constrained by energy conservation as it assumes all the material accreting through the outer thin disc.

 In order to constrain the thermal emission from the disk, we used the optical/UV fluxes in five bands measured with \swift{}. Thus, we fit the {\tt optxagnf} model modified by the Galactic X-ray absorption and interstellar optical/UV reddening jointly to the \swift{} UVOT and \suzaku{} XIS/PIN data. We used the {\scshape redden} model to account for the Galactic extinction ($E_{B-V} = 0.044$; \citep{1998ApJ...500..525S}). In {\tt optxagnf} model, the flux is determined by the four parameters: the black hole mass ($\rm M_{BH}$),  the spin of the black hole, the relative accretion rate ($\rm  L/L_{Edd}$), and the luminosity distance of the source ($\rm D_L$) so we fixed the norm to one. The black hole mass ($\rm M_{BH}$) was fixed  at $2.5 \times 10^{8}\msol$,~\citep{2010ApJ...717.1243G} and the luminosity distance of the source ($\rm D_L$) was fixed at $624 \mpc$. We varied the accretion rate ($\rm L/L_{Edd}$), $r_{corona}$ and the spin parameter $a^{*}$ to get the best fit. In XSPEC the model reads as {\tt wabs$\times$redden$\times$optxagnf}.

\begin{figure}

  \includegraphics[width=6.5cm,height=8cm,angle=-90]{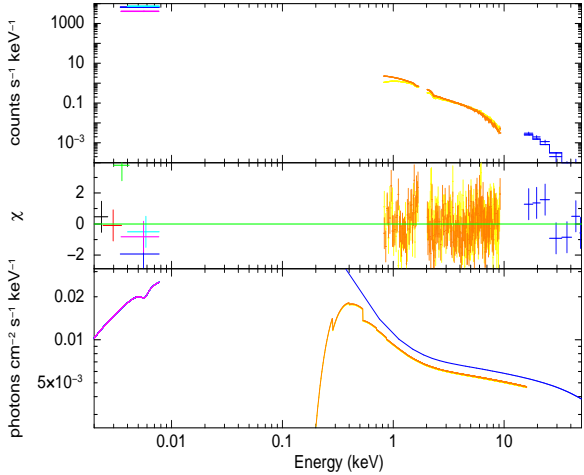}
  \caption{{\tt optxagnf} model fit of the combined \suzaku{} and \swift{}
    data for five combined spectra. Six points on the top left corner of
    the spectra represent \swift{} V, B, U, UW1, UW2 and UM2 data
    and the continuous yellow, orange and blue line represents XIS0 and XIS3 combined, XIS1 and PIN data
    respectively.}
\label{optxagn_combinedfit}
  
\end{figure}

\begin{figure*}
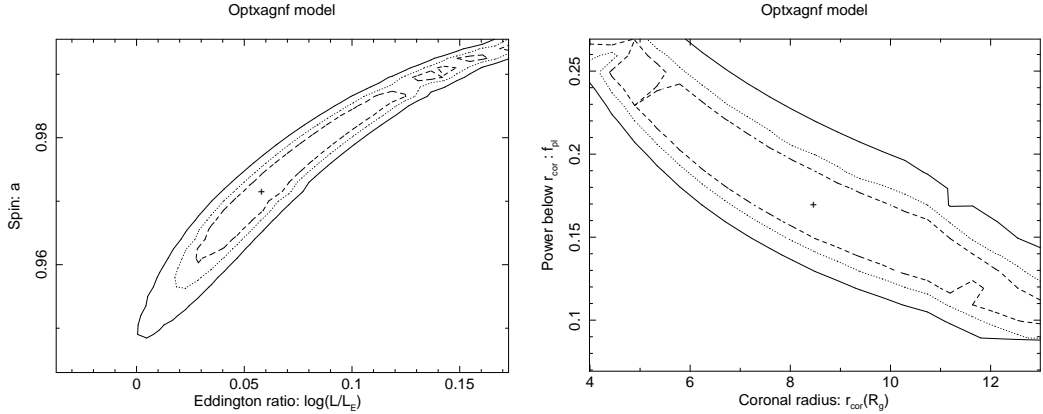


  \includegraphics[width=5.5cm,angle=-90]{contour_Le_vs_astar_BW_19thJuly_2015.ps}
  \includegraphics[width=5.5cm,angle=-90]{contour_Rcor_vs_Fpl_BW_20thJuly_2015.ps}
  \caption{{\it Left:} Contour plot of log of Eddington ratio vs spin parameter a and {\it Right:} Contour plot of coronal radius vs $f_{pl}$ for optxagnf best fit of five added spectra. Dashed, dotted and solid lines represent 68\%, 90\% and 95\% confidence contours respectively.}
\label{optxagnf_contour}
  
\end{figure*}

\begin{figure*}
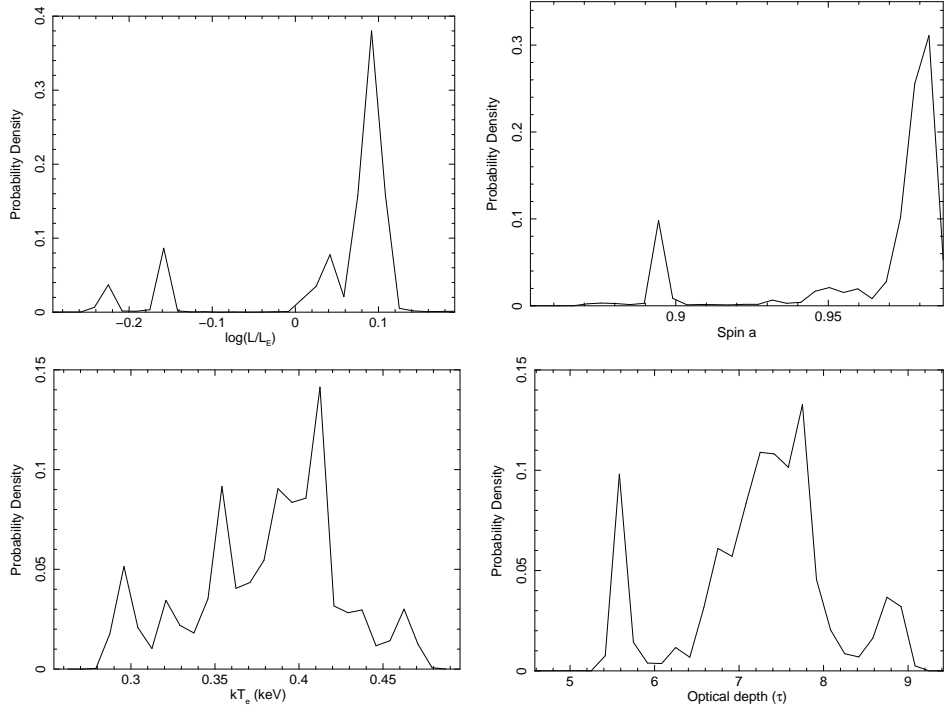


  \includegraphics[width=4.6cm,angle=-90]{Probability_distribution_logL_Le_20kchain.ps}
  \includegraphics[width=4.5cm,angle=-90]{Probability_distribution_spin_20kchain.ps}
  \includegraphics[width=4.7cm,angle=-90]{Probability_distribution_kte_20kchain.ps}
  \includegraphics[width=4.7cm,angle=-90]{Probability_distribution_tau_20kchain.ps}
 \caption{MCMC results:{\it Left top:} Probability distribution for $\log(\frac{L}{L_{E}})$, {\it Left bottom:} Probability distribution for $kT_{e}$, {\it Right top:} Probability distribution for $a$ and {\it Right bottom:} Probability distribution for $\tau$ for optxagnf best fit of five added spectra.}
\label{probability_optxagnf}
  
\end{figure*}

 First, we fitted those five spectra separately and found the best fit parameters to be consistent within errors(see Table~\ref{optxagnf fit}). Although the light-curves of the first three \suzaku{} observations show large variations in counr rates, the hardness ratio indicates that no significant variability is present during the five observation and the addition of five spectra together seem justified. So we combined the spectra of five data sets to minimize the variability effect of the UVOT data. The combined spectra produced a similar good fit $\chi^{2}$/dof=441/393 and are quoted in Table \ref{optxagnf fit}. The broadband spectral data of combined spectra and \swift{} UVOT, the folded {\tt optxagnf} model and the deviations are shown in Fig.\ref{optxagn_combinedfit}. We also tried to place a limit on the extinction of the host galaxy by adding a {\tt zredden} component to the optxagnf best fit and it reduced the $\dc \sim 5$ to $\chi^{2}$/dof=436/392 with the best fit value of $N_H = 1.73\times10^{20}\cmsqi$ and other parameters having values within the calculated error range from Table \ref{optxagnf fit}. In general the parameters of {\tt optxagnf} model are coupled with each other. To get an idea about the correlation of the best fit parameters we plotted the contours and are shown in~Fig.\ref{optxagnf_contour}. 
The probability density of the parameters, calculated following the MCMC fitting procedure, is shown in Fig.\ref{probability_optxagnf} and are consistent with the one from the {\tt relxill} best fit values (see~Fig.\ref{probability_relxill}). The best-fit parameters indicate towards a black hole accreting at almost Eddington or super-Eddington rate, which is in agreement with previous \xmm{} observations done by \citet{2007ApJ...656..691G}.
The Coronal radius (in $R_{g}=GM/c^{2}$) $\simeq 10$  marks the transition from colour temperature corrected blackbody emission to a Comptonised spectrum and importantly, the model resulted in the black hole spin parameter $a \sim 0.9$ for all five observations along with the combined spectra, which implies that the accretion disk extends very close to the black hole. The fit worsened drastically to $\chi^{2}$/dof=550/394 ($\dc \sim 108$), if we force the spin parameter to be zero. 

The \swift{}, optical points are not simultaneous with the \suzaku{} X-ray data and \swift{} XRT data was included to check the validity of {\tt optxagnf} best fit parameters and it provided a good fit with $\chi^{2}$/dof=78/81. The best fit parameter values (See Table.~{\ref{optxagnf fit}}) are in agreement with the values obtained using \suzaku{} data, although we note a significant drop in the lower limit of the Eddington rate ($L/L_{Edd}>0.4$) and also the spin parameter (a$>0.2$) and conclude that if only the {\tt optxagnf} model is correct to explain the observed spectra than the presence of a prograde black hole can be inferred.

\begin{table*}
\footnotesize
  \caption{Best fit parameters for \suzaku{}  and \swift{} observations of PKS~0558-504 for 
absorbed Optxagnf model\label{optxagnf fit}}
  \begin{tabular}{|l|l|llllll|l|} \hline
  & & & & Suzaku & & & & Swift \\\hline
 Component & Parameter & Obs-1 & Obs-2 & Obs-3 & Obs-4 & Obs-5 & Combined & XRT+UVOT \\ \hline


Gal. abs.  & $N_{H}(10^{20}{\rm cm^{-2}})$ & $ 4.4$ (f)             & $ 4.4$ (f)           & $ 4.4$ (f)         & $ 4.4$ (f)         &$ 4.4$ (f)   &$4.4$(f) &$4.4$(f) \\
Redden & E(B-V) & $0.044$ (f)& $ 0.044$ (f)        & $ 0.044$ (f)         & $ 0.044$ (f)       &$ 0.044$ (f)  &$ 0.044$ (f) &$ 0.044$ (f)\\ 	   
optxagnf   & $ M_{BH}{(10^8\rm~M\odot)}$           & $ 2.5$ (f)           & $ 2.5$ (f)         & $ 2.5$ (f)  &$ 2.5$ (f)  &$ 2.5$ (f) &$ 2.5$ (f) &$ 2.5$ (f)\\
           & $d{\rm~(Mpc}) $       & $ 624$ (f)              & $ 624$ (f)            & $ 624$ (f)           &$ 624$ (f)             &$ 624$ (f)  &$ 624$ (f) &$ 624$ (f)\\
          &  $\frac{L}{L_{E}}$          &$0.826^{+0.004}_{-0.004}$&$1.052^{+0.003}_{-0.001}$ & $0.858^{+0.001}_{-0.001}$ & $1.231^{+0.004}_{-0.002}$  &$0.928^{+0.004}_{-0.002}$ &$1.001^{+0.001}_{-0.001}$ &$0.476^{+0.543}_{-0.080}$\\ 
           &  $ kT_{e} (\kev)$       & $ >0.19 $                & $ >0.31 $               & $ >0.29 $             &$  >0.55 $              &  $ 0.29^{+0.19}_{-0.07} $  &$ >0.45    $ &$ 0.17^{+0.34}_{-0.03} $ \\ 
           &  $ \tau $               & $10.47^{+7.73}_{-2.34}$  & $9.07^{+3.49}_{-5.29}$  & $6.44^{+3.78}_{-3.51}$&$ 6.43^{+3.49}_{-5.89}$ &$11.44^{+4.49}_{-3.81}$ &$6.65^{+0.61}_{-0.25}  $ &$ >7.54    $ \\
           &  $ r_{cor}(r_{g})$      & $10.74^{+5.39}_{-4.91}$  &$10.83^{+5.16}_{-7.94}$  &$10.51^{+6.13}_{-8.07}$&$ 8.92^{+5.45}_{-3.98}$ &$10.70^{+1.81}_{-2.64}$ &$8.49^{+5.17}_{-3.76} $  &$ >9.07    $\\
           &  $ a $                  & $ 0.86^{+0.10}_{-0.39} $ & $ 0.94^{+0.05}_{-0.09}$ & $0.88^{+0.08}_{-0.04}$&$ 0.97^{+0.03}_{-0.03}$ &$0.91^{+0.06}_{-0.08}$  &$0.97^{+0.01}_{-0.37} $ &$0.50^{+0.39}_{-0.28} $\\
           &  $ f_{pl}$              & $0.26^{+0.17}_{-0.15}$   & $ 0.23^{+0.22}_{-0.10}$ & $0.20^{+0.23}_{-0.10}$&$ 0.19^{+0.14}_{-0.10}$ &$0.32^{+0.17}_{-0.14}$  &$0.17^{+0.30}_{-0.08} $ &$0.39^{+0.09}_{-0.13} $\\
           &  $ \Gamma $             & $2.12^{+0.07}_{-0.04}$   & $ 2.15^{+0.04}_{-0.09}$ & $2.02^{+0.07}_{-0.08}$&$ 2.08^{+0.05}_{-0.09}$ &$2.21^{+0.04}_{-0.04}$  &$2.13^{+0.02}_{-0.02}$ &$1.95^{+0.17}_{-0.21}$\\ \hline
           &  $\cd $                 & $833.48/729$           & $890.16/841$             & $698.93/671$          &$892.25/891$            & $883.57/873$ & $441.18/393$ & $78/81$\\ 
\hline \hline
\end{tabular} \\ 
Notes: (f) indicates a frozen parameter.
\end{table*}

 \subsection{Combined blurred reflection and intrinsic disk Comptonisation model \label{combined_model}}

\begin{figure}
  \centering
  \includegraphics[width=6.5cm,height=8cm,angle=-90]{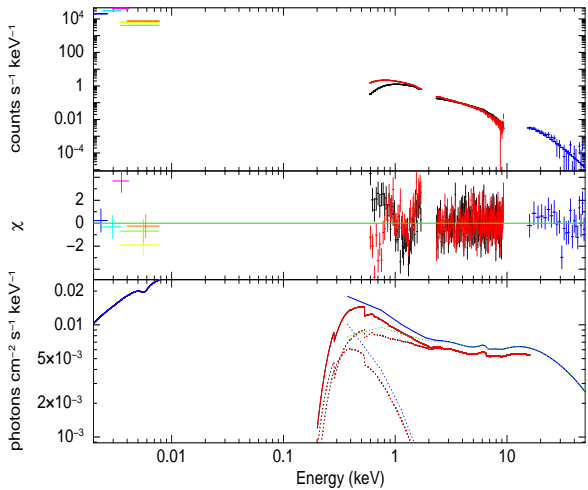}
  \caption{
Relxill fixed optxagnf model fit of the joint \suzaku{} and \swift{} data for five combined spectra. Six points on the top left corner represent \swift{} V, B, U, UW1, UW2 and UM2 data and the continuous black, red and blue line represent XIS0 and XIS3 combined, XIS1 and PIN data respectively.} 
\label{combined_model_fit}
  
\end{figure}

 It is possible that the soft X-ray excess is contributed by both the  blurred reflection and the intrinsic Comptonised disk emission. Using long \xmm{} observations, \citet{2010A&A...518A..28P} have shown that the UV and the soft excess exhibit a similar variability pattern and most likely arise in the accretion disk at distances of $\sim 40$ and $\sim 10r_g$. These results suggest that the soft excess is dominated by the intrinsic disk Comptonised emission though there may be a small contribution from the blurred reflection. Therefore, we have tried to model the broadband spectral data with a combination of the intrinsic Comptonised disk emission and relxill model.
We combined all the observations by adding the five sets of XIS and PIN data using the {\sc addascaspec} tool and obtained one set of XIS and PIN data.  As before, the blurred reflection and Comptonised disk emission provides satisfactory fit(see Table~\ref{Combined_fit_par}). To fit the combined reflection and disk Comptonised model, first we estimate the contribution of the blurred reflection to the soft excess emission by fitting the {\tt relxill} and 
powerlaw model to the spectral data above $2\kev$ (see Table~\ref{Combined_fit_par} for the best-fit parameters). Then we included the optical/UV data and replaced the $powerlaw$ model with {\tt optxagnf}. The {\tt relxill} model parameters were frozen to previously obtained best fit values. The best fit parameters of the combined model thus obtained are listed in Table~\ref{Combined_fit_par}. The combined spectral data, the best-fit folded blurred reflection plus Comptonised disk emission, the deviations and the unfolded model are shown in Fig.~\ref{combined_model_fit}.

\begin{table*}
\centering
\caption{Results of spectral fits to the \swift{} UVOT \& the combined \suzaku{} data set}
\label{Combined_fit_par}
\begin{tabular}{c c c c c c} \hline
\hline
 Model component & Parameter & Model: 1 & Model: 2 & Model: 3 & Model: 4\\
\hline 
\hline \\
Gal. abs.  & $N_{H}$($10^{20}{\rm~cm^{-2}}$) & $4.4$(f)       & $4.4$(f)    & $4.4$(f)      & $4.4$(f) \\

powerlaw   & $\Gamma $              & $2.37^{+0.01}_{-0.01}$  & -- & $2.16^{+0.19}_{-0.19}$ & -- \\
           & $n_{pl}(10^{-3})$      & $6.05^{+0.25}_{-0.77}$  & -- & $ <2.20$               & -- \\
relxill    &  $A_{Fe}$              & $4.43^{+0.56}_{-1.05}$  & -- & $0.89^{+1.09}_{-0.28}$ & $0.89$(f)\\ 
           &  $log\xi (\xiunit)$    & $2.97^{+0.05}_{-0.09} $ & -- & $3.17^{+0.14}_{-0.76}$ & $3.17$(f) \\ 
           & $ \Gamma $             & $2.37^{+0.01}_{-0.01}$  & -- & $2.16^{+0.19}_{-0.19}$ & $2.16$(f)\\
           &  $n_{rel}(10^{-5})$    & $0.76^{+0.05}_{-0.02}$  & -- & $0.74^{+0.46}_{-0.22}$ & $0.74$(f) \\
           &   $ q $                & $>9.81$                 & -- & $>9.40 $               & $9.98$(f)\\
           &   $ a $                & $>0.991$                & -- & $>0.987$               & $0.998$\\
           &   $ R_{in}(r_{g})$     & $1.27^{+0.02}_{-0.02}$  & -- & $1.43^{+0.15}_{-0.11}$ & $1.43$(f)  \\
           &   $ R_{br}(r_{g})$     & $2.89^{+0.06}_{-0.05}$  & -- & $2.93^{+0.06}_{-0.05}$ & $2.93$\\
           &   $ R_{out}(r_{g})$    & $400$(f)                & -- & $400$(f)               & $400$(f)\\
           &   $i(degree) $         & $45$(f)                 & -- & $45$(f)                & $45$(f)\\
Redden & E(B-V)                          & --    &$0.044$(f)  & -- &$ 0.044$(f)\\   
optxagnf   & $ M_{BH}{10^8\rm~(M\odot)}$ & --    &$2.50$(f)   & -- &$ 2.50$(f)   \\
           & $d{\rm~(Mpc}) $             & --    &$624$(f)    & -- &$ 624$(f)   \\
           &  $(\frac{L}{L_{E}})$        & --    &$1.001^{+0.001}_{-0.001}$     & -- &$ 0.83^{+0.33}_{-0.01}$  \\ 
           &  $ kT_{e} (\kev)$           & --    &$ >0.45    $                  & -- &$ 0.41^{+0.05}_{-0.05} $ \\ 
           &  $ \tau $                   & --    &$6.65^{+0.61}_{-0.25}  $      & -- &$ 8.48^{+0.95}_{-1.27}$ \\
           &  $ r_{cor}(r_{g})$          & --    &$8.49^{+5.17}_{-3.76} $       & -- &$ 9.63^{+6.85}_{-4.39} $ \\
           &  $ a $                      & --    &$0.97^{+0.01}_{-0.37} $       & -- &$ 0.79^{+0.07}_{-0.02}$ \\
           &  $ f_{pl}$                  & --    &$0.17^{+0.30}_{-0.08} $       & -- &$ <0.02$ \\
           &  $ \Gamma $                 & --    &$2.13^{+0.02}_{-0.02}$        & -- &$ 2.13$(f) \\
\hline
           &  $\cd $                  & $496.52/376$ & $441.18/393$& $259.28/292$ & $566.52/390$ \\ \hline
\end{tabular}

Note -- Model 1 : constant$\times$wabs$\times$(powerlaw$+$relxill);
Model 2 : constant$\times$wabs$\times$redden$\times$optxagnf;
Model 3 : constant$\times$wabs$\times$(powerlaw$+$relxill) above $2\kev$;
Model 4 : constant$\times$wabs$\times$redden$\times$(relxill$+$optxagnf);
 (f) indicates a frozen parameter.
\end{table*}

\section{Discussion \& Summary}

We performed detailed analysis of broadband optical/UV to hard X-ray spectrum of PKS~0558--504 using \suzaku{} and \swift{} observations.  These data allowed us to study the primary X-ray continuum, thermal Comptonisation, the optical/UV disk emission and the role of reflection in a radio-loud AGN. The observed X-ray emission is dominated by the primary X-ray continuum due to thermal Comptonisation of disk photons in a hot corona.
We confirmed the presence of prominent soft X-ray excess emission from PKS~0558--504 earlier detected by \citet{2010A&A...510A..65P}. The soft X-ray excess can be described either by blurred reflection or optically thick thermal Comptoisation. We estimate  spin of the black hole in this radio-loud AGN to be $a>0.2$ from the optical/UV disk emission and also assuming that the soft X-ray excess is the blurred reflection. Below we discuss implications of our results on the nature of the Comptonising hot corona, the inner extent of the accretion disk and black hole spin in the radio-loud AGN.



\subsection{The primary X-ray continuum and the Comptonising Hot Corona}

The broadband \suzaku{} spectral data allowed us to use physical Comptonisation models.
The high energy data are consistent with different geometries such as the spherical  or slab corona. In the Comptonised disk or the compps model, strong reflection is not required. This is consistent with the nondetection of broad iron line in the \suzaku{} data. 
We have measured electron temperature and optical depth of the hot corona to be $kT_e=256.5_{-13.6}^{+16.0}\kev$ and Compton y-parameter $y=0.09\pm0.01$ corresponding to optical depth $\tau\sim0.05$ (spherical corona) and $kT_e=93.3_{-17.2}^{+16.2}\kev$ and $y=0.38_{-0.08}^{+0.13}$ or $\tau\sim 0.5$ (slab corona). We notice the electron temperature in PKS~0558-504 is almost consistent with \nustar{} observations of other radio-loud AGNs e.g., in case of 3C~382 the $kT_e$ is found to be cooled down from $330\pm30\kev$ in the low flux data to $231^{+50}_{-88}\kev$ in the high
 flux data and the optical depth $\tau$  increased from 0.15 to 0.23~\citep{2014ApJ...794...62B}. 
For another radio-loud AGN 3C~273, $kT_e$ is found to have an upper limit of $320\kev$ and the optical depth $\tau = 0.15\pm0.8$~\citep{2015arXiv150606182M}. In case of Cygnus A the lower limit of the cutoff energy is measured to be $E_{cut}>111\kev$ at 90\% confidence and $>101\kev$ at 99\% confidence~\citep{2015arXiv150607175R}. The electron temperature is also higher and the optical depth is lower in PKS~0558--504 compared to that measured for the radio-quiet broad-line Seyfert 1 galaxy IC~4329A using long \nustar{} observations (spherical : $kT_e=33\pm6\kev$, $\tau=3.4_{-0.4}^{+0.6}$ and slab : $kT_e=37_{-6}^{+7}\kev$, $\tau=1.25_{-0.10}^{+0.20}$; \cite{2014ApJ...781...83B}). The electron temperature in the spherical geometry is also higher and the optical depth lower compared to that measured for the radio-quiet NLS1 SWIFT~J2127.4+5654 with \xmm{}$+$\nustar{} observations (spherical : $kT_e=53_{-26}^{+28}\kev$, $\tau=1.35_{-0.67}^{+1.03}$; slab : $kT_e=68_{-32}^{+37}\kev$, $\tau=0.35_{-0.19}^{+0.35}$; \cite{2014MNRAS.440.2347M}). These results suggest that the Comptonising corona in radio-loud AGN may be hotter compared to that of radio-quiet AGN. Using the optical/UV and X-ray data, we measure the disk luminosity ($L_{disk}\sim 2.6 \times 10^{45}\ergs$) and powerlaw luminosity ($\sim 5.5 \times 10^{44}\ergs$) which are the estimates of the accreted energy dissipated into the disk and the corona. Thus, about $20\%$ of the accreted energy is fed into the hot corona in PKS~0558--504.

\subsection{The accretion disk and black hole spin in radio-loud AGNs}
One of the major uncertainties in the study of the radio-loud AGN is the lack of knowledge of the inner extent of the accretion disk and its relation with the origin of the jet.
In AGN, the disk emission falls in the XUV band and it is not clear if the inner disk material or the Comptonising plasma or both these components are accelerated and ejected to form the jets. It has not been possible to determine both the existence and the inner extent of the accretion disks in radio-loud AGN. Though radio-observations have resolved the base of the jet and determined the size $\sim 5.5\pm0.4R_S$ for M~87. However, such observations do not reveal the inner extent of the disk and it is also not clear if the compact jet-base is really confined within $6R_S$ or it is located further away, but projected along our line of sight. We have tried to estimate the inner extent of the disk in the radio-loud NLSy-1 PKS~055--504 using two ways -- ($i$) from the optical/UV emission, assuming it to arise from the accretion disk and ($ii$) from the soft excess, assuming it to be the blurred reflection from the disk. 

In the intrinsic disk Comptonisation model, accretion rate relative to the Eddington rate through the outer disk is calculated from the optical/UV luminosity. The total luminosity observed over the entire optical to X-ray band is given by $L_{total} = \eta \dot{M} c^2$ where the emissivity depends on the black hole spin (see ~\citep{2012MNRAS.420.1848D}). The black hole mass of PKS~0558--504 estimated from four different techniques ranges from $6.7\times10^7{\rm~M_{\odot}}$ to $3\times10^{8}{\rm~M_{\odot}}$ with a mean of $1.8\times10^{8}{\rm~M_{\odot}}$ \citep{2010ApJ...717.1243G}. Using the mean black hole mass, the intrinsic disk Comptonised model provided the spin parameter $a\geq0.5$ for all \suzaku{} observations) and $a=0.5_{-0.3}^{+0.4}$ for the \swift{} UVOT+XRT data alone (see Table~\ref{optxagnf fit}. This suggests the possibility of an accretion disk extending very close to the black hole. 


In the blurred reflection model, the amount of relativistic blurring  required to produce the observed smooth, soft X-ray excess component determines the inner extent of the accretion disk. This model provided satisfactory fits to all the five \suzaku{} broadband spectra and resulted in very small inner disk radius $<3r_g$ (see Table~\ref{relxill fit}). The non-detection of iron line could be due to the extreme width of the line that may arise from the innermost region as suggested by high emissivity index and small inner radius. Now, in general the chi-square space is complicated for these models and it is appropriate to make sure the fit is not in a local minimum. The MCMC technique is used to search the high-dimensional parameter space and evaluate the uncertainties on model parameters and the fitting procedure yielded a similar probability density as has been obtained for the chi squared fitting and the lack of large excursions or trends in results(Fig~\ref{probability_relxill} and Fig~\ref{probability_optxagnf}) suggests that we have indeed found the global best fit of these spectral models. Thus, both the blurred reflection and Comptonised disk model suggest a spinning black hole in PKS~0558--504 to fit the spectra and considering no direct evidence for relativistic reflection except the soft excess it is justified to say that if the blurred reflection or the optxagnf model is correct to predict the soft excess, then a moderate black hole spin can be inferred or at least we can impose a model-dependent lower-bound($a>0.2$) on the spin.

Again recent findings of broad iron L line and soft X-ray lags at high frequency imply significant contribution of the blurred reflection to the soft X-ray excess emission. At the same time, soft X-ray leads at low frequencies imply that the soft X-ray excess cannot  entirely be the blurred reflection. Based on the observation of soft X-ray leads on long time scales in the NLS1 Akn~564, \citet{2007ApJ...671.1284D} concluded that that the soft excess cannot be the reprocessed emission and had suggested a compact, cool corona responsible for the soft excess. Recently, \citet{2014MNRAS.442.2456G} have modeled the lag spectra, as a combination of lags due to propagating fluctuations in the accretion flow resulting soft leads on long-time scales and reverberation lags on short time scales producing the soft lags. In this model, fluctuations start in the disk and propagate to an inner optically thick, low temperature region responsible for the soft X-ray excess via thermal Comptonisation, and then into the innermost hot corona responsible for the hard powerlaw component. The soft leads and lags require contribution from both the optically thick thermal Comptonisation and blurred reflection. In such a scenario, the inner disk radius inferred from either the blurred reflection  or the intrinsic disk model alone can be misleading. In Sect.~\ref{combined_model}, we showed that the combination of blurred reflection and the intrinsic Comptonised disk model fitted to the combined broadband \suzaku{} data result in high black hole spin ($a=0.98_{-0.08}^{+0.01}$). 

 We note that our estimates of the inner extent of the accretion disk and the black hole spin are dependent on the models for the soft excess. While the origin of the soft X-ray excess is still ambiguous in nature, there are only two physical models currently available. The blurred reflection model most frequently results in high black hole spin for NLS1 galaxies due to the smoothness of the soft excess, and the optically thick thermal Comptonisation model has no spin parameter. However, when the optical/UV emission is combined with the soft excess emission,  the $optxagnf$ models also resulted in spinning black hole in PKS~0558--504. 
%
Thus, we conclude that the presently available physical models i.e. blurred reflection for the soft excess and the  intrinsic thermal Comptonisation model for the optical/UV and soft excess, are able to explain the origin of soft X-ray excess in this radio-loud AGN and our results suggest that the disk truncation at large radii and retrograde black hole spin both seem unlikely to be the necessary conditions  for launching the jets.

\section{Acknowledgements}
We thank the anonymous referee for many useful comments and suggestions which helped to improve the quality of the manuscript. RG acknowledges the financial support from  Visva-Bharati University and IUCAA visitor programme. RG, thanks Sibasish Laha for helpful discussions. This research has made use of archival data of \suzaku{} and \swift{} observatories through the High Energy Astrophysics Science Archive Research Center Online Service, provided by the NASA Goddard Space Flight Center.


\bibliographystyle{mnras} \bibliography{mybib}

\end{document}